\documentclass[preprint]{aastex}
\usepackage{emulateapj5}
\usepackage{apjfonts}
\usepackage{natbib}
\def\gtorder{\mathrel{\raise.3ex\hbox{$>$}\mkern-14mu
             \lower0.6ex\hbox{$\sim$}}}
\def\ltorder{\mathrel{\raise.3ex\hbox{$<$}\mkern-14mu
             \lower0.6ex\hbox{$\sim$}}}
\def\ltsima{$\; \buildrel < \over \sim \;$}
\def\simlt{\lower.5ex\hbox{\ltsima}}
\def\gtsima{$\; \buildrel > \over \sim \;$}
\def\simgt{\lower.5ex\hbox{\gtsima}}
\begin{document}
\submitted{}
\title{Comment on ``Measuring the black hole masses of high redshift quasars''}

   \author{Dan Maoz}

 \affil{School of Physics \& Astronomy and Wise Observatory,
    Tel-Aviv University, Tel-Aviv 69978, Israel. dani@wise.tau.ac.il}

\begin{abstract}
In a recent paper, McLure \& Jarvis (2002, astro-ph/0204473, v.1) reanalyze
AGN broad-line reverberation measurements 
presented in Kaspi et al. (2000). They find the broad-line region size
and the AGN luminosity at 5100~\AA\ can be related by
$R_{BLR}\propto L_{5100}^{0.5}$. This differs from the result of Kaspi et al.,
who found $R_{BLR}\propto L_{5100}^{0.7}$, a departure from expectations
for a constant ionization parameter. In this note I investigate the source
of the discrepancy. McLure \& Jarvis use quasar luminosities based
on the single-epoch, multichannel-spectrophotometer measurements of Neugebauer
et al. (1987), obtained in 1980. I show that the Neugebauer et al. fluxes
are systematically higher, by a constant flux offset, compared with the
multi-epoch CCD measurements of Kaspi et al., obtained concurrently with
the echo-mapping radii. In addition, McLure \& Jarvis erred when converting
these fluxes to luminosities. The two effects compound to a typical
factor 2 (0.3 dex)
overestimate of the luminosities of the PG quasars in the sample. Since
McLure \& Jarvis did adopt the Seyfert luminosities in Kaspi et al. (2000),
they obtained a slope that is flatter by 0.2.
 
\end{abstract}

\section{Introduction}

Reverberation mapping of active galactic nuclei (AGNs) exploits the 
light-travel-time delay between continuum variations and the response
of broad emission lines to measure the sizes of AGN broad-line regions
(BLRs). This size can then be used, together with the estimated
velocities of the BLR gas to derive masses for the central black holes
(See Netzer \& Peterson 1996, for a review.) The size-luminosity and
mass luminosity relations of AGNs may shed new light on understanding
these objects and the connections between the black holes in AGNs
and those found in normal nearby galaxies.

Kaspi et al. (2000) measured reverberation sizes, $R_{BLR}$, for 17 PG quasars
between 1991--1998 using the Wise Observatory 1m telescope and the 
Steward Observatory 2.3m telescope, with a combination of CCD
spectrophotometry and photometry. The time averaged flux for each
quasar was derived from 20-70 epochs per quasar. The rest-frame
5100 \AA\ continuum flux densities
 were measured from the $\sim 10$ \AA\ resolution spectra
while minding potential systematics such as broad-emission-line wings
and atmospheric absorptions in the higher-redshift objects.
Following Galactic extinction corrections (small for most of the objects)
the absolute fluxes were used to calculate 5100 \AA\ luminosities, 
$\lambda L_{\lambda}(5100 {\rm \AA})$. The sizes
and luminosities of Seyfert galaxies with reverberation measurements
were compiled from previously published works.
 A regression analysis taking into account the errors
in both $R_{BLR}$ and $\lambda L_{\lambda}(5100{\rm \AA})$ showed that 
$R_{BLR} \propto [\lambda L_{\lambda}(5100{\rm \AA})]^{0.70\pm 0.03}$. 
This was a surprising
result, since it has long been speculated that the dependence would be to
the power 0.5. Such a scaling would lead to an ionization parameter
(ratio of ionizing photon density to electron density) at the surface of
the BLR clouds that is independent of luminosity, and would explain the similarity
of AGN spectra over many orders of magnitude in luminosity.

In a recent paper, McLure \& Jarvis (2002) examine to what degree UV 
observables, namely 3000\ \AA\ luminosities and Mg~II line velocities, can 
serve the purpose of the optical observables -- 5100 \AA\ luminosities
and H$\beta$ widths -- used to date in AGN black hole estimates.
In the course of their work, they re-analyze the data presented by
Kaspi et al. (2000) and conclude that actually
$R_{BLR} \propto [\lambda L_{\lambda}(5100{\rm \AA})]^{0.50\pm0.02}$ is the  best fit,
contrary to the result of Kaspi et al., and
consistent with the expectations for a constant ionization parameter.

Vestergaard (2002) has carried out an analysis along similar lines.
After studying the various systematics that can affect the slope 
determination, she concludes that the best estimate (her equation A5) is
$R_{BLR} \propto [\lambda L_{\lambda}(5100)]^{0.66\pm0.09}$, consistent
with the result of Kaspi et al. (2000). The statement by McLure \& Jarvis
(2002), that Vestergaard (2002) found a slope consistent with 0.5 
but decided to
adopt a slope of 0.7 anyway, is incorrect.
 
In this Note, I investigate the source of the discrepancy between
Kaspi et al. and McLure \& Jarvis. I show that it arises, first, due to the
adoption by McLure \& Jarvis of old, single-epoch, and systematically 
higher fluxes, but only
for the high luminosity part of the sample; and second, due to incorrect
conversion from flux to luminosity. This is confirmed and corrected in 
a revised version of their paper (R. McLure, private communication).

\section{Analysis}

McLure \& Jarvis chose to base their
luminosities for the PG quasars in the Kaspi et al. (2000) sample 
on the measurements by Neugebauer et al. (1987). These measurements were
obtained in 1980 with a multichannel spectrophotometer mounted on the
Palomar 5m telescope. The measurements had coarse resolution of $\sim
300$ \AA, making the avoidance of emission and absorption lines in the
measurement more difficult. In the numbers tabulated by Neugebauer et al.,
interpolation is needed, for most of the quasars, to obtain the flux at
rest wavelength 5100~\AA. As opposed to the measurements of Kaspi et al. 
(2000), only one epoch per object exists, inducing scatter due to the time
variability of quasars. More importantly, the Neugebauer et al. (1987)
measurements were obtained about 20 years before the Kaspi et al. (2000)
size measurements, and it has been shown in Seyfert galaxies that BLR
size and emission-line width change with time in individual objects
(Peterson \& Wandel 2000).

Figure 1 compares the 
observer-frame quasar flux densities measured at wavelengths corresponding
to rest-wavelength 5100\ \AA, as reported by Kaspi et al. (2000),
to the corresponding numbers obtained from Neugebauer et al. (1987), 
after the necessary interpolation and conversion to $f_{\lambda}$ in 
the latter dataset. Apart from the large differences in a few objects,
presumably due to variability, there is a clear additive systematic 
offset between 
the datasets, in the sense that the Neugebauer et al. (1987) measurements
have higher flux. While one cannot say for sure which dataset is at fault,
it is possible that the Palomar data suffer from imperfect
sky  subtraction -- a task that is much easier with present-day two-dimesional
CCD detectors. To get a ``third opinion'' I have compared the Kaspi et al. 
(2000) flux densities to those indicated by the data of
 Boroson \& Green (1992). The Boroson
\& Green (1992) fluxes are generally {\it lower} than the Kaspi et al. fluxes
by about 30\%. This difference is reasonable, given that they were obtained
through a narrow $1.5''$ slit, and at non-parallactic angles, and hence
some light loss is expected.

\vspace{-1in}
\hspace{-1in}
{\includegraphics[angle=0,width=6in]
{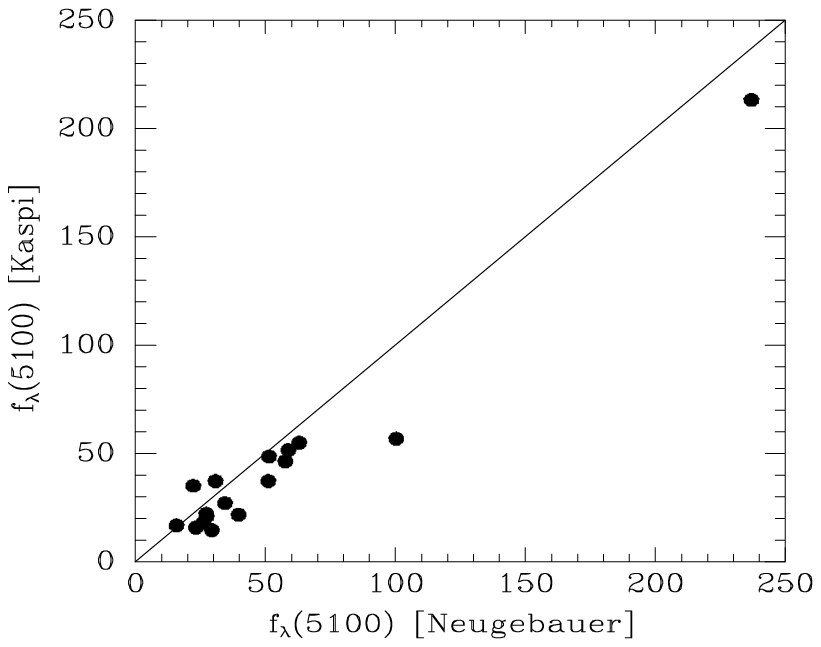}}
\begin{quote}
\vspace{-.3in}
\baselineskip3pt
{\footnotesize Fig.~1-- 
Comparison of $f_{\lambda}[5100 {\rm \AA} (1+z)]$
flux densities for PG quasars measured by Kaspi et al. (2000)
and by Neugebauer et al. (1987). Units are $10^{-16}$ erg s$^{-1}$ cm$^{-2}$
 \AA$^{-1}$.

\vspace{0.1in}
Fig.~2-- 
Comparison of luminosities $\lambda L_{\lambda}$
at rest wavelength 5100 \AA, calculated from each of the datasets
in Fig. 1.  
}
\end{quote}

\vspace{-.9in}
\hspace{-1in}
{\includegraphics[angle=0,width=6in]
{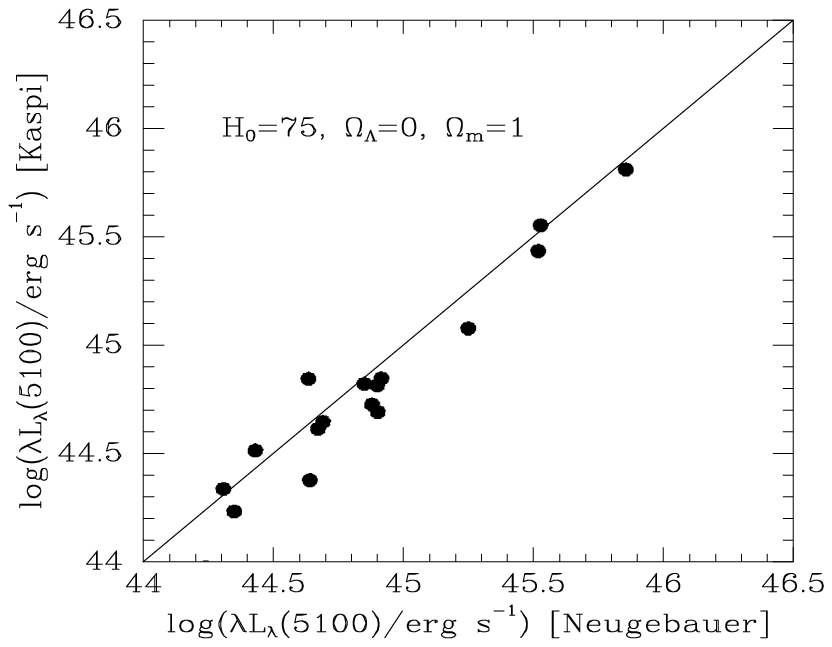}}
\vspace{-.75in}

\vspace{1.2in}

Figure 2 examines the effect of this flux discrepancy on the calculated
luminosities. It compares the 5100 \AA\ luminosities I have calculated
from the Neugebauer et al. (1987) fluxes to the luminosities found
by Kaspi et al. (2000). In both cases, the cosmology assumed is
$H_0=75$, $\Omega_{m}=1$, and $\Omega_{\Lambda}=0$. As expected, the lower
luminosity quasars are most affected, and their luminosities are about
0.1 dex higher when based on the Neugebauer et al. (1987) numbers.
This conclusion was also reached by Vestergaard (2002), and Figure 2 is
nearly identical to her Figure 2a. Indeed, comparing directly her and my
numbers for the luminosities calculated from the Neugebauer et al. data, 
I find only small differences of order 5\%, which are 
ascribable to the interpolation of the
Neugebauer et al. data.

In Figure 3, I compare the quasar luminosities listed by  McLure \& Jarvis,
to my own calculation based on the same Neugebauer et al. data, assuming
the cosmology cited by McLure \& Jarvis, $H_0=70$, $\Omega_{m}=0.3$, 
and $\Omega_{\Lambda}=0.7$. Specifically, I calculate
$$
L_{\lambda}(5100{\rm \AA})=4\pi d_L^2 (1+z) f_{\lambda}[5100{\rm \AA} (1+z)],
$$
where, for the flat ($\Omega_{total}=1$)
cosmologies we consider here, the luminosity distance is
$$
d_L=c H_0^{-1} (1+z)\int_0^z[\Omega_m (1+z)^3+\Omega_{\Lambda}]^{-0.5} dz.
$$
Figure 3 indicates that there is an error in McLure \& Jarvis's conversion to 
luminosities, resulting in too-high luminosities by about 0.2 dex,
especially at the higher luminosities. R. McLure (private communication) 
confirms that this is a programming error that resulted in an extra factor
$(1+z)^2$.

\vspace{-0.1in}
\hspace{-1in}
{\includegraphics[angle=0,width=6in]
{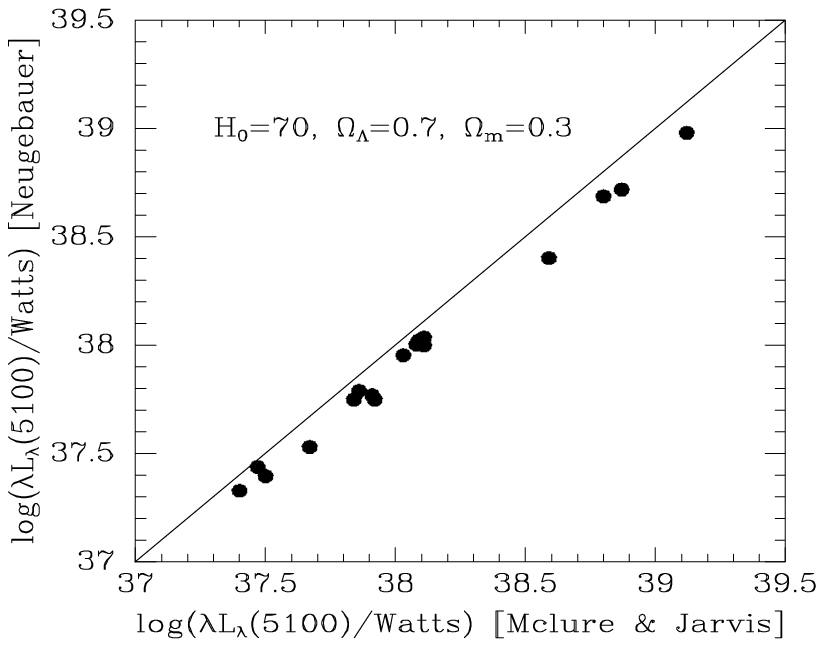}}
\begin{quote}
\baselineskip3pt
{\footnotesize Fig.~3-- 
Comparison of 5100 \AA\ luminosities reported
by McLure \& Jarvis (2002) with luminosities calculated here from
the same Neugebauer et al. (1987) data they used, and for the same 
cosmology.
}
\end{quote}

\newpage

Finally, in Figure 4, I compare the quasar luminosities
listed by McLure \& Jarvis, to the luminosities of Kaspi et al. (2000),
after converting the latter to the cosmology and units used by the former.
Here one sees the compound effect of the higher fluxes and the
incorrect conversion to luminosity used by McLure \& Jarvis.
This leads to a total offset of about 0.3 dex (a factor 2). 

\vspace{0in}
\hspace{.1in}
\centerline{\includegraphics[angle=0,width=6in]
{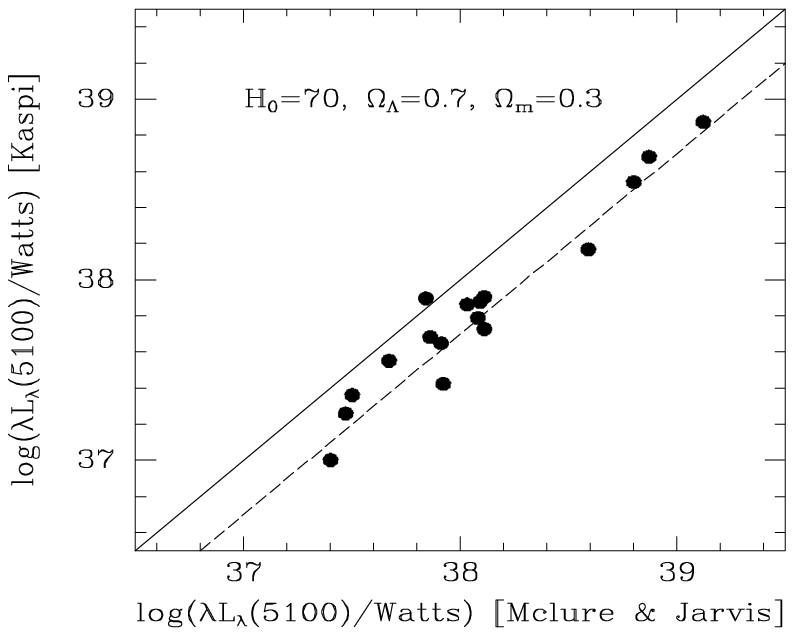}}
\begin{quote}
\vspace{-0.5in}
\baselineskip3pt
{\footnotesize Fig.~4-- 
Comparison of the Kaspi et al. (2000) luminosities
and the McLure \& Jarvis (2002) luminosities, after converting
the former to the cosmological model and units used by the latter.
}
\end{quote}

The systematically higher, by 0.3 dex, quasar luminosities used by 
McLure \& Jarvis, can explain the flatter slope they obtained
in the $R_{BLR}$ vs  $L_{\lambda}(5100 {\rm \AA})$ relation. While preferring 
the Neugebauer et al. measurements over those of Kaspi et al. for the
quasars, McLure \& Jarvis adopted the Seyfert luminosities as
given in  Kaspi et al. (2000). Since the Seyfert $R_{BLR}$ and $L_{\lambda}(5100 {\rm \AA} )$ 
values are, respectively, about 0.8 dex and  1.15 dex lower 
than those
of the quasars, increasing the quasar luminosities by 0.3 dex, will lower
the slope of the relation from 0.7 to  0.55. Some additional flattening
also arises from the use of the $\Omega_{\Lambda}$-dominated cosmology,
which makes the higher-redshift quasars in the sample slightly more luminous.
In a revised version of their paper (R. McLure private communication),
 McLure \& Jarvis find a best-fit slope of 0.6 when using the Neugebauer et al.(1987) data,
and $0.65\pm 0.11$ when using the Kaspi et al. fluxes, fully consistent with
Kaspi et al. (2000) and Vestergaard (2002). 

\acknowledgements

I thank Shai Kaspi for stimulating discussions and for suggesting and
performing the comparison to Boroson \& Green (1992). Ross McLure is thanked
for helping sort out the problem.

\end{document}